\documentclass{article}

\usepackage{arxiv}

\usepackage{amsmath,amssymb,amsfonts}
\usepackage{algorithmic}
\usepackage{graphicx}
\usepackage{bbm}
\usepackage{textcomp}
\usepackage{algorithmic}
\usepackage{algorithm}
\usepackage{booktabs}
\usepackage{multirow}
\usepackage{xspace}
\usepackage{url}

\usepackage[utf8]{inputenc} % allow utf-8 input
\usepackage[T1]{fontenc}    % use 8-bit T1 fonts
\usepackage{hyperref}       % hyperlinks
\usepackage{url}            % simple URL typesetting
\usepackage{booktabs}       % professional-quality tables
\usepackage{amsfonts}       % blackboard math symbols
\usepackage{nicefrac}       % compact symbols for 1/2, etc.
\usepackage{microtype}      % microtypography
\usepackage{lipsum}		% Can be removed after putting your text content
\usepackage{graphicx}
\usepackage{natbib}
\usepackage{doi}

\usepackage{hyperref}

\newcommand\FOREACH[1]{\FOR{\textbf{each} #1}}
\def\BibTeX{{\rm B\kern-.05em{\sc i\kern-.025em b}\kern-.08em T\kern-.1667em\lower.7ex\hbox{E}\kern-.125emX}}

\title{Defending Against Adversarial Attack in ECG Classification with Adversarial Distillation Training}

\author{\hspace{1mm}Jiahao Shao \\
    Department of Industrial Engineering\\
	Tsinghua University\\
	Beijing, China 100084 \\
	\texttt{shaojh19@mails.tsinghua.edu.cn} \\
	\And
	\hspace{1mm}Shijia Geng \\
	HeartVoice Medical Technology\\
	Hefei, China 230088\\
	\texttt{gengshijia@heartvoice.com.cn} \\
	\And
	\hspace{1mm}Zhaoji Fu \\
	School of Management\\
	University of Science and Technology of China\\
	Hefei, China 230026,\\
	and HeartVoice Medical Technology\\
	Hefei, China 230088\\
	\texttt{fuzj@mail.ustc.edu.cn} \\
	\And
	\hspace{1mm}Weilun Xu \\
	HeartRhythm Medical\\
	Beijing, China 100020\\
	\texttt{xuweilun@heartrhythm.cn} \\
	\And
	\hspace{1mm}Tong Liu \\
	Department of Cardiology\\
	Tianjin Institute of Cardiology\\
	Second Hospital of Tianjin Medical University\\
	Tianjin, China 300210\\
	\texttt{liutongdoc@126.com} \\
	\And
	\hspace{1mm}Shenda Hong \thanks{Corresponding author.}\\
	National Institute of Health Data Science\\
	Peking University,\\
	and Institute of Medical Technology\\
	Health Science Center of Peking University\\
	Beijing, China 100191\\
	\texttt{hongshenda@pku.edu.cn} \\
}
	
\renewcommand{\shorttitle}{DEFENDING AGAINST ADVERSARIAL ATTACK IN ECG CLASSIFICATION}

\hypersetup{
pdftitle={Defending Against Adversarial Attack in ECG Classification with Adversarial Distillation Training},
pdfsubject={cs.AI},
pdfauthor={Jiahao Shao, Shijia Geng, Zhaoji Fu, Weilun Xu, Tong Liu, Shenda Hong},
pdfkeywords={deep learning, electrocardiograms, adversarial training, distillation, adversarial attack},
}

\begin{document}
\maketitle

\begin{abstract}
In clinics, doctors rely on electrocardiograms (ECGs) to assess severe cardiac disorders. Owing to the development of technology and the increase in health awareness, ECG signals are currently obtained by using medical and commercial devices. Deep neural networks (DNNs) can be used to analyze these signals because of their high accuracy rate. However, researchers have found that adversarial attacks can significantly reduce the accuracy of DNNs. Studies have been conducted to defend ECG-based DNNs against traditional adversarial attacks, such as projected gradient descent (PGD), and smooth adversarial perturbation (SAP) which targets ECG classification; however, to the best of our knowledge, no study has completely explored the defense against adversarial attacks targeting ECG classification. Thus, we did different experiments to explore the effects of defense methods against white-box adversarial attack and black-box adversarial attack targeting ECG classification, and we found that some common defense methods performed well against these attacks. Besides, we proposed a new defense method called Adversarial Distillation Training (ADT) which comes from defensive distillation and can effectively improve the generalization performance of DNNs. The results show that our method performed more effectively against adversarial attacks targeting on ECG classification than the other baseline methods, namely, adversarial training, defensive distillation, Jacob regularization, and noise-to-signal ratio regularization. Furthermore, we found that our method performed better against PGD attacks with low noise levels, which means that our method has stronger robustness.
\end{abstract}

% keywords can be removed

\section{Introduction}

Electrocardiograms (ECGs) are widely used by clinicians to diagnose a range of cardiovascular diseases, which are the leading cause of death worldwide \cite{cardiovascular}. Owing to the development of technology and the increase in people's awareness to health, many companies have developed wearable devices that can measure single-lead ECG signals, such as the Huawei Watch GT2 Pro ECG and Apple Watch Series 4, which are worn by millions of people. Using these wearable devices, people can detect whether they have cardiovascular diseases before the disease becomes severe. However, it is impossible for clinicians to spend a considerable amount of time analyzing the large amount of ECG signals collected by these devices.

Deep neural networks (DNNs) are an economic alternative approach for classifying multi-lead ECG signals\cite{Shensheng2018Towards,2020A} and single-lead ECG signals \cite{2020Non}. In addition, owing to the development of this technology, the accuracy of DNNs is comparable to that of professional cardiologists \cite{hannun2019cardiologist,sinnecker2020deep,elul2021meeting}. DNNs have been successfully used in many ECG analysis tasks \cite{hong2020opportunities,somani2021deep}, such as cardiovascular management \cite{siontis2021artificial,fu2021artificial},  disease detection \cite{attia2019artificial,erdenebayar2019deep,raghunath2020prediction,ribeiro2020automatic,hong2020holmes}, sleep staging \cite{banluesombatkul2020metasleeplearner}, biometric human identification \cite{labati2019deep,hong2020cardioid}, and ECG-based non-invasive monitoring of blood glucose \cite{2021Non}, indicating the effectiveness of DNNs in ECG analysis \cite{hughes2021performance}.

\begin{figure}
\centering
	\hspace{-0.5cm}
    \includegraphics[scale = 0.6]{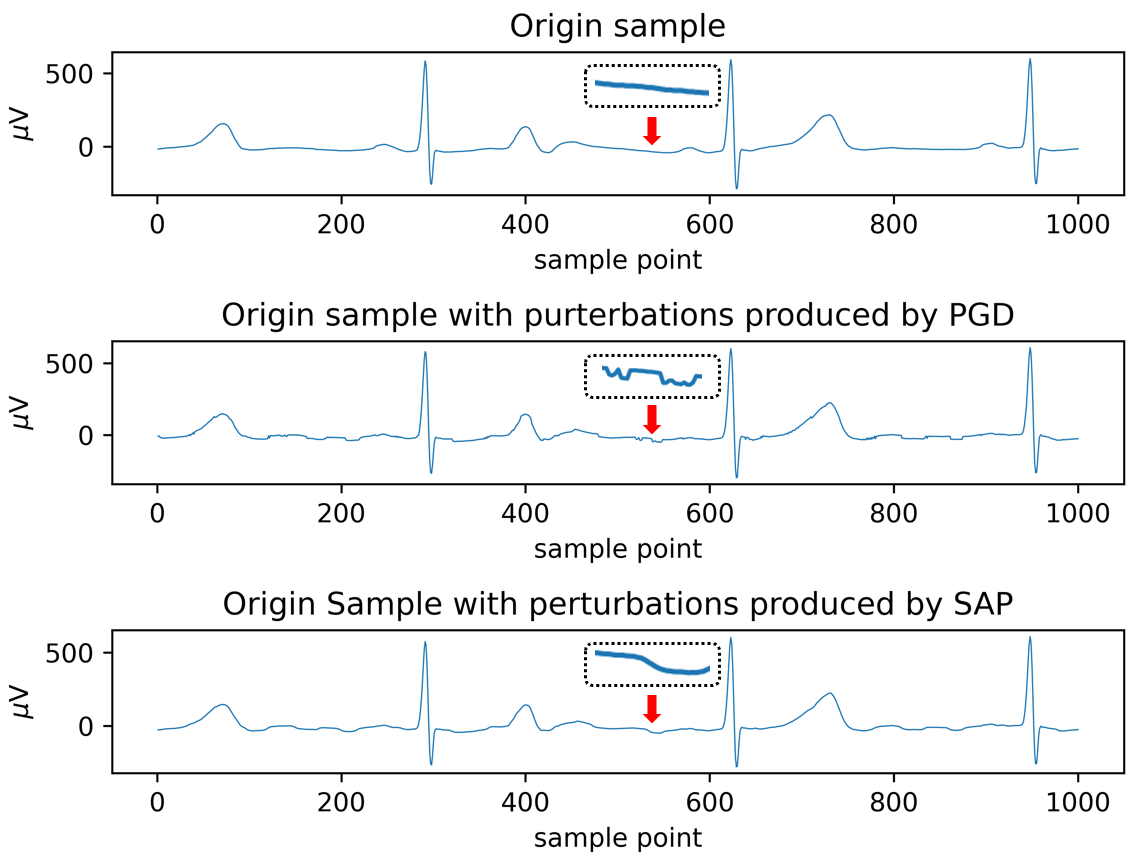}
    \caption{Comparison Between an Original ECG signal and That Attacked by PGD and SAP}
    \label{fig:differe}
\end{figure}

%begin
However, DNNs are vulnerable when facing adversarial noises involving perturbations that are imperceptible to the human eye. This phenomenon was first discovered by Szegedy et al.\cite{2013Intriguing} in the image classification field. Subsequently, researchers proposed certain convenient methods for generating adversarial perturbations, such as fast gradient sign method\cite{goodfellow2015explaining}, basic iterative method\cite{2016Adversarial}, projected gradient descent (PGD)\cite{2017Towards}, and Carlini and Wagner (C\&W) attacks\cite{carlini2017evaluating}. These methods are mainly aimed at attacking DNNs for image classification, and they cannot be extended directly to DNNs for ECG signals, because the perturbations created by these methods are not physiologically plausible\cite{2020Deep}. To attack DNNs for ECG signal classification, several white-box and black-box adversarial attack methods have been proposed recently. The white-box adversarial attack is generated by utilizing the inner structure knowledge of the target DNN, whereas the black-box adversarial attack does not have any knowledge regarding the network's inner structure. The white-box attack methods proposed by Han et al.\cite{2020Deep} and Chen et al.\cite{2019ECGadv} are similar to PGD and C\&W attacks, respectively. The only difference is that the perturbations created by smooth adversarial perturbation (SAP) proposed by Han et al. are smoothed through convolution, whereas those created by the attack method of Chen et al. are significantly limited by setting up an objective function to maximize the smoothness of the attack. Detecting the perturbations becomes difficult because of the restriction of the objective function or convolution processing. Lam et al.\cite{10.1145/3417312.3431827} proposed a black-box attack called boundary attack, which improves the smoothness of perturbations by using a low-pass Hanning filter. In \autoref{fig:differe}, we plot a part of an original ECG signal sample and its counterparts that are attacked by PGD and SAP. We can see that the signal attacked by PGD is unnatural and not physiologically plausible, but it is difficult to distinguish the signal attacked by SAP from natural ECG signals.

To defend against adversarial attacks in ECG signal classification, Yang et al.\cite{9313442} applied the gradient-free trained sign activation neural network to classify ECG signals and found that the perturbations created by the HopSkipJump boundary-based black-box attack can fool the classification network and are visually distinguishable. Furthermore, because the network is gradient-free and white-box attacks mainly use gradient information to create adversarial attacks, the network is immune to traditional white-box attacks. Ma and Liang \cite{ma2020enhance} explored the effectiveness of three defense methods against PGD and SAP attacks, namely, adversarial training, Jacobian regularization (JR), and noise-to-signal ratio (NSR) regularization. The results showed that all three methods can improve the robustness of the DNNs for ECG classification against PGD and SAP attacks, and NSR has the best performance among these defense methods. However, both Yang et al. and Linhai et al. didn't completely explore the defense against the white-box and black-box adversarial attacks. In addition, the accuracy of the gradient-free trained sign activation network proposed by Yang et al.\cite{9313442} is lower than that of traditional DNNs on certain data, and it can be achieved or surpassed by traditional DNNs using certain defense methods.

In this study, we explored the defense against the white-box and black-box adversarial attacks which are aimed at ECG-based DNNs. Furthermore, SAP \cite{2020Deep} is applied to represent the white-box attack and boundary attack \cite{10.1145/3417312.3431827} is applied to represent the black-box attack. We defended ECG-based DNNs against SAP and boundary attack with common defense methods, such as adversarial training, defensive distillation, JR and NSR regularization, and found these methods performed well against SAP and boundary attack. Furthermore, defensive distillation can learn class-related knowledge, and transfer the knowledge from the first network to the second network to generalize the classification ability. While SAP and boundary attack make adversarial ECG samples by adding small perturbations into original ECG samples, so that DNNs classify adversarial ECG samples into wrong categories, and those perturbations are so small that it is difficult for people to distinguish those adversarial samples. Therefore, we can regard those small perturbations as reasonable fluctuations of ECG signals, and add the ECG samples with those small perturbations into the training set of distillation network, which can make distillation network to further learn fluctuations of ECG signals and class-related knowledge to improve generalization ability. Based on this idea, we proposed a new method called Adversarial Distillation Training (\textbf{ADT}) in which we added adversarial samples into the training process of defensive distillation. The results show that ADT outperforms JR, NSR regularization, defensive distillation and adversarial training under SAP and boundary attack.

Furthermore, although perturbations created by traditional adversarial attacks, such as PGD attacks, are not physiologically plausible, if we keep the level of noise, $\epsilon$, low for PGD attacks, we will obtain fewer unnatural ECG segments. It is possible for hackers to attack ECG signals collected by wearable devices with low-noise PGD attacks because clinicians do not check these signals and people in general may not be able to recognize the attacked patterns. In addition, We can consider low-noise PGD attacks as a robust test for ADT. Thus, we use the trained DNNs with different defense methods to classify ECG signals attacked by low-noise PGD attacks, and the results show that ADT still performs well against this type of attack.

\begin{figure}
\centering
    \includegraphics[scale = 0.8]{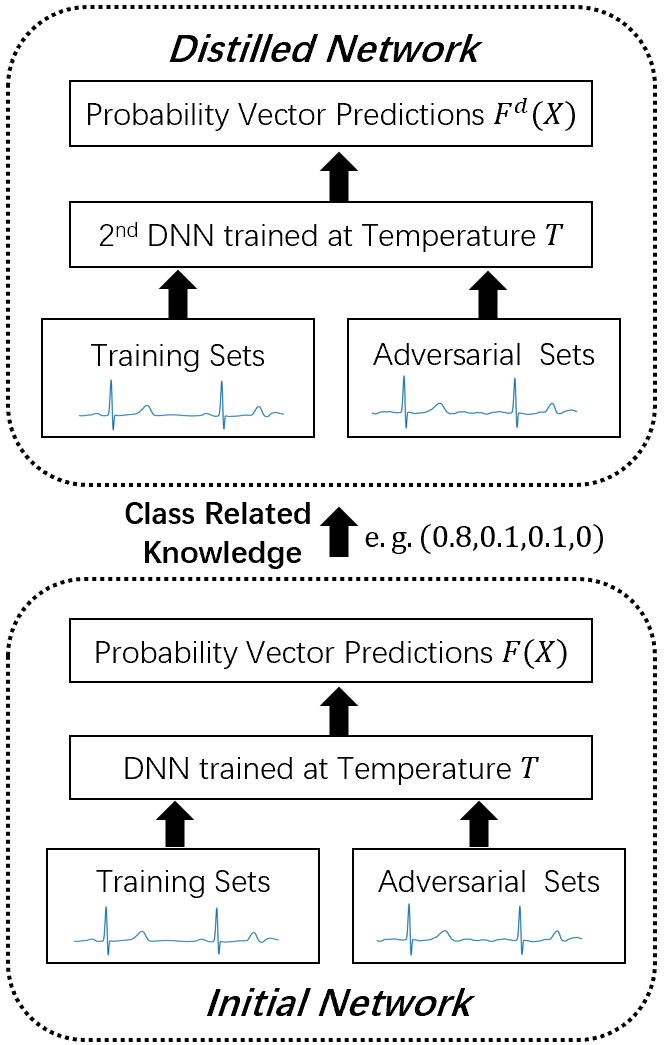}
    \caption{Training Process of Adversarial Distillation Training.}
    \label{fig:process}
\end{figure}

\section{Method}

In this section, we will discuss the details of defensive distillation and our method, ADT.

\subsection{Defensive Distillation}

Initially, distillation learning was used exclusively to reduce the hierarchy of DNNs\cite{2015Distilling}. Generally, a large-scale DNN is first trained to learn the distribution of the sample data, and then, the labels of the training samples are changed to the probability of each category for the samples predicted by the large-scale DNN. Subsequently, a small-scale DNN network is trained with the changed training data without the loss of accuracy. The idea of this method is that the parameters learned by the DNN can represent the characteristics of data and the probability vector output from the softmax layer of the network contains some knowledge of the data. For example, if the probability values corresponding to two categories of samples predicted by a DNN are similar, it means that there are some similarities between these two types of data. Later, Nicolas et al.\cite{2016Distillation} proposed defensive distillation, in which the initial network is the same as the distilled network.

The core of defensive distillation is the temperature parameter, $T$, which is the only additional parameter compared with the ordinary DNN and is used in the normalization process of the softmax layer as follows.
\begin{equation}\label{equ:soft}
F_{i}(X) =\frac{e^{z_{i}(X)/T}}{\sum_{l=0}^{N-1}e^{z_{l}(X)/T}}, i \in 0,1,...,N-1
\end{equation}

In the equation, $z_{i}(X)$ denotes the output logit of the last layer of the DNN for sample $X$ corresponding to category $i$. For convenience, we use $z_{i}$ to represent $z_{i}(X)$. Furthermore, $Z(X) = (z_{0},z_{1},...,z_{N-1})$ denotes the output logit vector. The softmax layer normalizes the output logit vector, $Z(X)$, through \autoref{equ:soft}, and the result value, $F_{i}(X)$, denotes the probability that sample $X$ belongs to category $i$. Here, we have $F(X) = (F_{0}(X), F_{1}(X),..., F_{N-1}(X))$ as the output probability vector of the softmax layer for the DNN. We can see that the larger the $T$ value is, the smaller the difference between the values of the probability vector, $F(X)$, becomes, and as $T \to \infty$, $F_{i}(X)$ converges to $\frac{1}{N}$. As $T = 1$, \autoref{equ:soft} is the same as that of the traditional DNN.

Here, we use $Y(X)$, an indicator vector, to denote the labels of sample $X$, with the non-zero element in $Y(X)$ representing the correct category, and if $X$ belongs to the first category, $Y(X)$ is like $(e.g., (1,0,0,...,0))$. Under defensive distillation, we need to train our classification DNN model twice, and both times, the model must be trained from the beginning. For the first time, the loss function is as follows:
\begin{equation}\label{equ:loss1}
-\frac{1}{|\chi|}\sum\limits_{X \in \chi}\sum\limits_{i \in (0,...,N-1)}Y_{i}(X)\log F_{i}(X),
\end{equation}

where $\chi$ denotes the entire training set, and the goal is to minimize the loss function. Because $Y_{i}(X)$ has only one non-zero value, 1, \autoref{equ:loss1} can be changed as
\begin{equation}\label{equ:loss1n}
-\frac{1}{|\chi|}\sum\limits_{X \in \chi}\log F_{l}(X),
\end{equation}

where $l$ is the index of the correct label for sample $X$. For the second training, the labels for the training set are changed as the output probability vectors of the first trained DNN, which are called soft labels. The loss function is
\begin{equation}\label{equ:loss2}
-\frac{1}{|\chi|}\sum\limits_{X \in \chi}\sum\limits_{i \in (0,...,N-1)}F_{i}(X)\log F^{d}_{i}(X).
\end{equation}

The goal of the training is to minimize the loss function. Based on the study conducted by Nicolas et al.\cite{2016Distillation}, we have the following.
\begin{align*}\label{equ:diff}
    \frac{\partial F_{i}(X)}{\partial X_j}\Big|_{T} &=\frac{\partial}{\partial X_j}\left(\frac{e^{z_{i}/T}}{\sum_{k=0}^{N-1}e^{z_{k}/T}}\right)  \\
    &=\frac{e^{z_{i}/T}}{Tg^{2}(X)}\left(\sum_{k=0}^{N-1}\left(\frac{\partial z_{i}}{\partial X_j} -\frac{\partial z_{k}}{\partial X_j}\right)e^{z_{k}/T}\right)
\end{align*}

This means that with an increase in $T$, the elements of the Jacobian matrix of $F$ decrease. In other words, the gradients of our classification DNN decrease. Because gradients are used in SAP to create perturbations, it is more challenging for an attacker to create successful noises to fool the classification DNN as the gradients decrease. Therefore, defensive distillation makes our classification DNN model insensitive to small changes of the input ECG signals as the parameter $T$ increases.

On the other side, for the training process of the first network of defensive distillation, the optimization mechanism is adjusting the parameter of the network to make $F(X)$ converge to $Y(X)$. This kind of training mechanism often makes DNNs with only one network over fit the training data. However, defensive distillation applies $F(X)$ output by the first network as the soft labels of training data for the second network, and the goal of optimization mechanism will make $F^{d}(X)$ converge to $F(X)$, but $F^{d}(X)$ can't be the same as $F(X)$ in practice, which improves the generalization ability of the second network.

\subsection{Adversarial Distillation Training}
However, due to the limitation and contingency of nature ECG signals, it is difficult to learn class-related knowledge and data fluctuation well with only nature training data. Thus, we need to add some adversarial ECG samples created by SAP into the training process of defensive distillation, and we call the new adversarial defense method as Adversarial Distillation Training (ADT).

Studies have shown that learning the characteristics of adversarial samples which is called adversarial training improves the robustness of the classification model. Szegedy et al.\cite{2013Intriguing} first found that DNNs are vulnerable to adversarial perturbations, and they used adversarial training to improve the robustness of the DNN. They found that it was better to use adversarial perturbations in the hidden layer. However, Goodfellow et al.\cite{goodfellow2015explaining} discovered that if the activation function of the hidden layer for the neural network is unbounded, such as the ReLU function, adding adversarial perturbations to the inputs is better. While it is time-consuming with adversarial samples created by PGD added into the training process of DNNs. To solve this problem, Shafahi et al.\cite{2019Adversarial} proposed a new training algorithm in which the gradient information is recycled to update the parameters of the network. In other words, every time the gradients of the adversarial samples are calculated, the network parameters are updated according to the gradients, and this process is repeated. In contrast, in traditional adversarial training, the network parameters are updated after the final calculation of the gradients of adversarial samples. In our study, the last perturbation generated by SAP is smoother than that generated in the intermediate process, and this can make the DNN find its real shortcomings. Thus, we use only the last perturbation generated by SAP instead of the large fluctuation one generated in the intermediate process. Subsequently, we update the parameters of the DNN after the final calculation of the gradients of adversarial samples. Certain new adversarial training algorithms exist, such as max-margin adversarial training\cite{ding2020mma} and increasing-margin adversarial training\cite{ma2021increasingmargin}, which are suitable for defending large perturbations. However, the noise level of SAP is not high, indicating that these new adversarial training methods are not suitable for defending SAP; therefore, we do not consider these new training algorithms.

The training process with adversarial samples as training data can be regarded as an optimization problem\cite{2017Towards}, and the form is as follows.
\begin{equation}
\min \limits_{\theta} E_{(x,y) \sim D}\left[\max \limits_{\delta} L(\theta, x + \delta, y) \right]
\end{equation}
The inner function describes the process of creating adversarial samples with $\delta$ representing the adversarial perturbations, and the target is to maximize the inner loss function. The outer function denotes the training process with adversarial samples, and the goal is to minimize the loss function by adjusting the value of the neural network parameters, $\theta$. In this training process, the training set consists of only adversarial samples, which lack the training of the original samples. Therefore, we adopt the strategy with a mixture of adversarial samples and original samples as training data, and this is shown as follows.
\begin{equation}\label{equ:mixedloss}
\min \limits_{\theta} \left(c E_{(x,y)}\left[L_{adv}(\theta, x_{adv}, y) \right] + (1-c)E_{(x,y)}\left[L(\theta, x, y) \right] \right)
\end{equation}
Where $x_{adv}$ denotes adversarial sample. In this way, we not only ensure the accuracy of the trained model on the natural samples, but also defend against attacks of adversarial samples by learning the characteristics of these two types of samples.

For adversarial attacks, if a hacker makes a classifier mistakenly identify a sample as a specified category, it is called a target attack. If the hacker makes the model classify a sample mistakenly without specifying the label of the error classification, it is called a non-target attack. In this study, we applied non-target SAP attack to create adversarial samples.

Furthermore, the first step to create adversarial samples for SAP is to create traditional adversarial samples using PGD. Generally, PGD creates adversarial samples by using multiple iterations and limits the difference between new adversarial samples and those created in the last iteration. We use $Clip_{x,\epsilon}(x')$ to represent limiting the maximum difference between $x$ and $x'$ to $\epsilon$, where $\epsilon$ denotes the noise level, and the larger $\epsilon$ is, the greater the fluctuation of noise. We first set $x'_{0}=x$; then, we have
\begin{equation}\label{equ:pgd}
x'_{i} = Clip_{x'_{i-1},\epsilon}(x'_{i-1} + \alpha sign(\bigtriangledown_{x'_{i-1}}L(f(x'_{i-1},y))))
\end{equation}
After $t$ iterations, traditional adversarial samples are created, and $x_{adv}=x'_{t}$. We define $\delta$ as the adversarial perturbation, which is the difference between the adversarial sample and the corresponding original sample. Then, SAP makes $\delta$ smooth through convolution, which is expressed as follows.
\begin{equation}\label{equ:conv}
x_{adv}(\delta) = x + \frac{1}{m}\sum_{i}^{m}\delta \otimes K(s[i],\sigma[i])
\end{equation}
In \autoref{equ:conv}, $K(s[i],\sigma[i])$ denotes a Gaussian kernel of size $s[i]$ and standard deviation $\sigma[i]$. Next, replacing s[i] with 2M+1 and simplifying $K(s[i],\sigma[i])$ as K with $\sigma[i]$ as $\sigma$, we have
\begin{equation}\label{equ:setk}
(\delta \otimes K)[n] = \sum_{m=1}^{2M+1}\delta[n-m+M+1] \times K[m]
\end{equation}
and $K[m]$ is
\begin{equation}\label{equ:Km}
K[m] = \frac{exp(-\frac{(m-M-1)^2}{2\sigma^2})}{\sum_{i=1}^{2M+1}exp(-\frac{(i-M-1)^2}{2\sigma^2})}
\end{equation}
SAP uses a process similar to PGD to update perturbations, $\delta$, by maximizing the loss function of the classification DNN. Similarly, we set $\delta'_{0}=\delta$, and we have
\begin{equation}\label{equ:updc}
\delta'_{i} = Clip_{\delta'_{i-1},\epsilon}(\delta'_{i-1} + \alpha sign(\bigtriangledown_{\delta'_{i-1}}L(f(x_{adv}(\delta'_{i-1}),y))))
\end{equation}
After $t'$ steps, we obtain the final adversarial permutation, $\delta'_{t'}$, and the adversarial sample, $x_{adv} = x + \delta'_{t'}$.

In ADT, we not only add adversarial samples created by SAP into the training process of the first network of ADT to learn the class-related knowledge and data fluctuations, but also add them into the training process of the second network of ADT to further improve the generalization ability of the classification model. The entire training process of the proposed method is shown in \autoref{fig:process}, and the entire process is detailed in \autoref{alg:joint}.
\begin{algorithm}
\caption{The Training Process of Adversarial Distillation Training}
\label{alg:joint}
\begin{algorithmic}[1]
    \REQUIRE Training set $(X_D,Y_D)$
    \FOR{epoch = 1 $\rightarrow$ $E1$}
        \FOREACH{mini-batch $(X,Y)$ of $(X_D,Y_D)$}
            \STATE Create adversarial samples $(X_{adv},Y)$ through \autoref{equ:pgd}-\autoref{equ:updc};
            \STATE Calculate initial DNN output logits of $X$ and $X_{adv}$;
            \STATE Calculate probability of each category of $X$ and $X_{adv}$ through \autoref{equ:soft};
            \STATE Calculate loss through \autoref{equ:loss1} and mixed loss through \autoref{equ:mixedloss};
            \STATE According to the mixed loss, calculate gradients of initial DNN parameters;
            \STATE Update initial DNN parameters;
        \ENDFOR
    \ENDFOR
    \STATE Calculate the soft labels of $X_D$, $Y'_{D}$, generate new training set $(X_D, Y'_{D})$
    \FOR{epoch = 1 $\rightarrow$ $E2$}
        \FOREACH{mini-batch $(X,Y)$ of $(X_D,Y'_{D})$}
            \STATE Create adversarial samples $(X_{adv},Y'_{D})$ through \autoref{equ:pgd}-\autoref{equ:updc};
            \STATE Calculate distilled DNN output logits of $X$ and $X_{adv}$;
            \STATE Calculate probability of each category of $X$ and $X_{adv}$ through \autoref{equ:soft};
            \STATE Calculate loss through \autoref{equ:loss1n} and mixed loss through \autoref{equ:mixedloss};
            \STATE According to the mixed loss, calculate gradients of distilled DNN parameters;
            \STATE Update distilled DNN parameters;
        \ENDFOR
    \ENDFOR

\end{algorithmic}
\end{algorithm}

\section{Experiments}

\subsection{Datasets}
In our experiments, the data are obtained from the publicly available training dataset of the 2017 PhysioNet$/$CinC Challenge\cite{2017AF}. All these ECG signals are single-lead, and their lengths are approximately 9--61 s. The available dataset contains 8528 ECG signals, including 5076 normal ECGs, 758 atrial fibrillation (AF) ECGs, 2415 other ECGs, and 279 noise samples. It is obvious that there is a severe category imbalance problem in the data, and to solve the problem, we duplicate the noise samples five times and double the size of the AF ECG samples.

Because the length of the data is not fixed and it is not suitable for the training of the classification network, we first limit the length of the ECG signals to 9000 sampling points. For ECG signals with less than 9000 sampling points, we fill the same number of zeros on both sides of the data. For ECG signals with more than 9000 sampling points, we only take the first 9000 data points. During the process of training the DNN, the dataset after data expansion and length limitation is divided into two parts: 90\% for the training set and the remaining 10\% for the test set.

\subsection{Compared Method}
To demonstrate the effectiveness of our proposed model, we used several adversarial defense methods for comparison.

\textbf{Compared Methods Group 1: baselines.} We apply JR\cite{2018Improving}, NSR regularization\cite{2020Improve}, defensive distillation and adversarial training as four baseline methods for comparison. In particular, JR penalizes large gradients with respect to the input. It adds the square of gradients to the loss function, which limits the fluctuation of gradients with respect to the input noise. By comparison, NSR penalizes significant changes in output logits with respect to small changes in input. It adds the ratio of the change in logits due to the noise of input to the original logits in the loss function. Based on the studies conducted by Ma and Liang\cite{ma2020enhance}, the only parameter of JR, $\lambda$, is set as 44, due to its outstanding performance, and the two parameters of NSR regularization, $\epsilon_{max}$ and $\beta$, are set to 1. To make the classification model converge quickly, the regularization term of JR and NSR regularization and the NSR margin loss are not added to the training process until the 11th epoch. If we do not add any adversarial sample into the training process of both networks of ADT, the training mechanism is the same as defensive distillation. If we only use one network and add adversarial samples into the training process of the network, it is the same as the mechanism of adversarial training. Thus, we apply defensive distillation and adversarial training as other two baseline methods, and we denote defensive distillation as \textbf{DD} and adversarial training as \textbf{AT}.

\textbf{Compared Methods Group 2: variants of our method.} In our method, we add adversarial samples into the training process of both networks of ADT. It is interesting to investigate adding adversarial samples to only one of the two networks during the training process. Thus, we applied these two variants of our method as the other two comparison methods. Furthermore, we abbreviate the two methods as \textbf{Init-ADT} and \textbf{Dist-ADT}, respectively, where \textbf{Init-ADT} denotes the methods with adversarial samples in the training process of the first network of defensive distillation and \textbf{Dist-ADT} denotes the methods with adversarial samples in the training process of the second network of defensive distillation.

Notably, we do not develop different classification DNNs for different defense methods. Our classification model is always a 13-layer convolution network\cite{2018Towards}, and these defense methods are only special settings to enhance the robustness of the classification DNN.

\subsection{Evaluations}
In this study, we used the accuracy ratio and F1-score as performance metrics. Specifically, the accuracy ratio is calculated by dividing the number of truly classified samples by the total number of samples, and the F1-score is the harmonic mean of the F1-score from the classification type. \autoref{table:f1value} lists the counting rules for the numbers of different variables. The F1-score for each category of the ECG signal is defined as follows.

Normal: $F_{1N}=\frac{2 \times Nn}{\sum{N}+\sum{n}}$,

AF: $F_{1A}=\frac{2 \times An}{\sum{A}+\sum{a}}$,

Other: $F_{1O}=\frac{2 \times Oo}{\sum{O}+\sum{o}}$,

Noise: $F_{1P}=\frac{2 \times Pp}{\sum{P}+\sum{p}}$. 

Here, the F1-score is calculated as
\begin{equation}
    F_{1}=\frac{F_{1N}+F_{1A}+F_{1O}+F_{1P}}{4}
\end{equation}

\begin{table}
\centering
\caption{Evaluations of experiments. }\label{table:f1value}
% \resizebox{0.7\columnwidth}{!}{%
\begin{tabular}{l|llllll}
\hline
                              & \multicolumn{6}{c}{Prediction}               \\
\hline
                              &        & Normal   & AF   & Other   & Noise   & Total \\
\multirow{5}{*}{Ground-truth} & Normal & $Nn$     & $Na$ & $No$    & $Np$    & $\sum{N}$     \\
                              & AF     & $An$     & $Aa$ & $Ao$    & $Ap$    & $\sum{A}$     \\
                              & Other  & $On$     & $Oa$ & $Oo$    & $Op$    & $\sum{O}$     \\
                              & Noise  & $Pn$     & $Pa$ & $Po$    & $Pp$    & $\sum{P}$     \\
                              & Total  & $\sum{n}$ & $\sum{a}$  & $\sum{o}$     & $\sum{p}$     &  $\sum{All}$     \\
\hline
\end{tabular}
% }
\label{table:eval}
\end{table}

\subsection{Implementation Details}

Here, we introduce the experimental implementation details. For the base deep model, we applied the 13-layer convolution network \cite{2018Towards}, which is one of the top-tier models in the 2017 PhysioNet/CinC Challenge.

The entire training process of ADT is shown as \autoref{alg:joint}, and in the experiment, we set $E1=E2=100$. The parameters $s$ and $\sigma$ of the Gaussian kernel are set as $\{5,7,11,15,19\}$ and $\{1,3,5,7,10\}$, respectively. In the process of creating adversarial samples, the iteration step to create PGD adversarial samples, $t$, is set to $5$, and the iteration step to smooth the adversarial perturbation, $t'$, is also set to $5$. In addition, we set $c$ in \autoref{equ:mixedloss} as $0.5$ and $\alpha$ in both \autoref{equ:pgd} and \autoref{equ:conv} as 1. For defensive distillation, the training epochs of the initial network and distilled network are also set as $100$. For all other defense methods, the number of training epochs was $100$. We set the training batch size of all models to $16$ and applied the Adam optimizer with 0.001 as the initial learning rate.

In the process of creating adversarial samples to attack defense models, many parameters are kept the same as those in the training of our method, except that $t$ and $t'$ are set as variable parameters, so that we can know about the defense effect of those methods under different smoothing-degree sample attacks. In addition, we used a fixed test dataset to create adversarial samples. Furthermore, where the value of T is not described, it defaults to 1.

Moreover, to determine the accurate defense effects of different methods against SAP and PGD attacks, and avoid the results deviation caused by randomness, we train the classification DNN with each baseline defense method and ADT five times.

\section{Results}
\subsection{Defense Effects against SAP Attacks}
Here, we first attacked original test samples by SAP($t=20$, $t'=40$) based on trained DNNs without defense methods, and then, we used each trained DNN with defense methods to classify these attacked test samples. \autoref{table:results_1} shows the average prediction accuracy, F1-score, decline ratio and corresponding standard deviation of five trained DNNs with each defense method in this situation. However, the attacker may know what defense methods we use, then train a similar model and attack it. Therefore, we are curious about the robustness of the trained DNNs with defense methods when the attacker uses themselves to make adversarial samples. In this paper, the former situation is called situation I, and the latter situation is called situation II. \autoref{table:results1} shows the performance of the trained DNNs with defense methods in situation II.

\autoref{table:results_1} and \autoref{table:results1} show that under SAP attacks, the accuracy of the classification DNN is significantly reduced (a decrease of about $50\%$) and the F1-score is reduced to $28\%$. From \autoref{table:results_1}, we can see that all defense methods performed well in situation I, even JR which performed worst can control the decline ratio at $11\%$, and keep the accuracy ratio and F1-score at $0.7676$, $0.6772$ respectively. While in situation II, the performance of some defense methods was not good, including JR and DD, and the trained DNNs with JR even had a worse performance than the trained DNNs without any defense method. In these two situations, our proposed method, ADT at $T=1$, had the best defense performance, and it controlled the decline of model accuracy within the range of less than $1\%$ in situation I and $5\%$ in situation II. Besides, ADT at $T=1$ also had a good performance in F1-score, and it even increased the F1-score of the classification model a little in situation I.

Furthermore, we can see that NSR regularization performs better than JR under SAP attacks and can maintain the accuracy of the classification model above $80\%$ in situation I and above $70\%$ in situation II, indicating that the defense effect of NSR is good. In our experiments, adversarial training exhibits outstanding performance, making the classification model maintain an accuracy ratio above $80\%$ and an F1-score of more than $60\%$ in both situations under SAP attacks. In addition, the performances of the methods that add adversarial samples into the training process of only one network of defensive distillation are excellent, such as Dist-ADT and Init-ADT. It is easy to understand that Dist-ADT has a better defense effect than Init-ADT: adversarial samples are added into the training process of the second network of DD for Dist-ADT, and it is the second network that is used to classify ECG samples, whereas Init-ADT puts adversarial samples in the first network of DD, which does not take the task of classify data. DD behaved well in situation I and had a similar performance to JR. While it did not perform as well as JR in situation II, with only $50.79\%$ accuracy ratios and $33.97\%$ F1-scores, which denotes that DD and JR could defend against adversarial samples targeted at the trained DNNs without any defense method, but they couldn't defend against those adversarial samples targeted at the trained DNNs with themselves, that is to say, their robustness is not strong; on the contrary, other defense methods have good robustness.

\begin{figure*}
\centering
\begin{tabular}{cc}
\multicolumn{2}{c}{\includegraphics[width=0.8\linewidth]{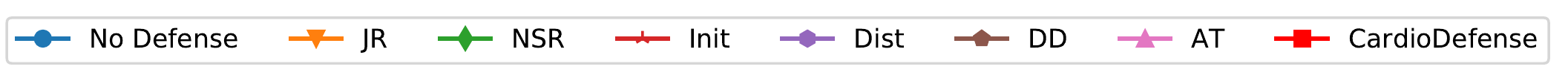}} \\
\includegraphics[width=0.45\linewidth]{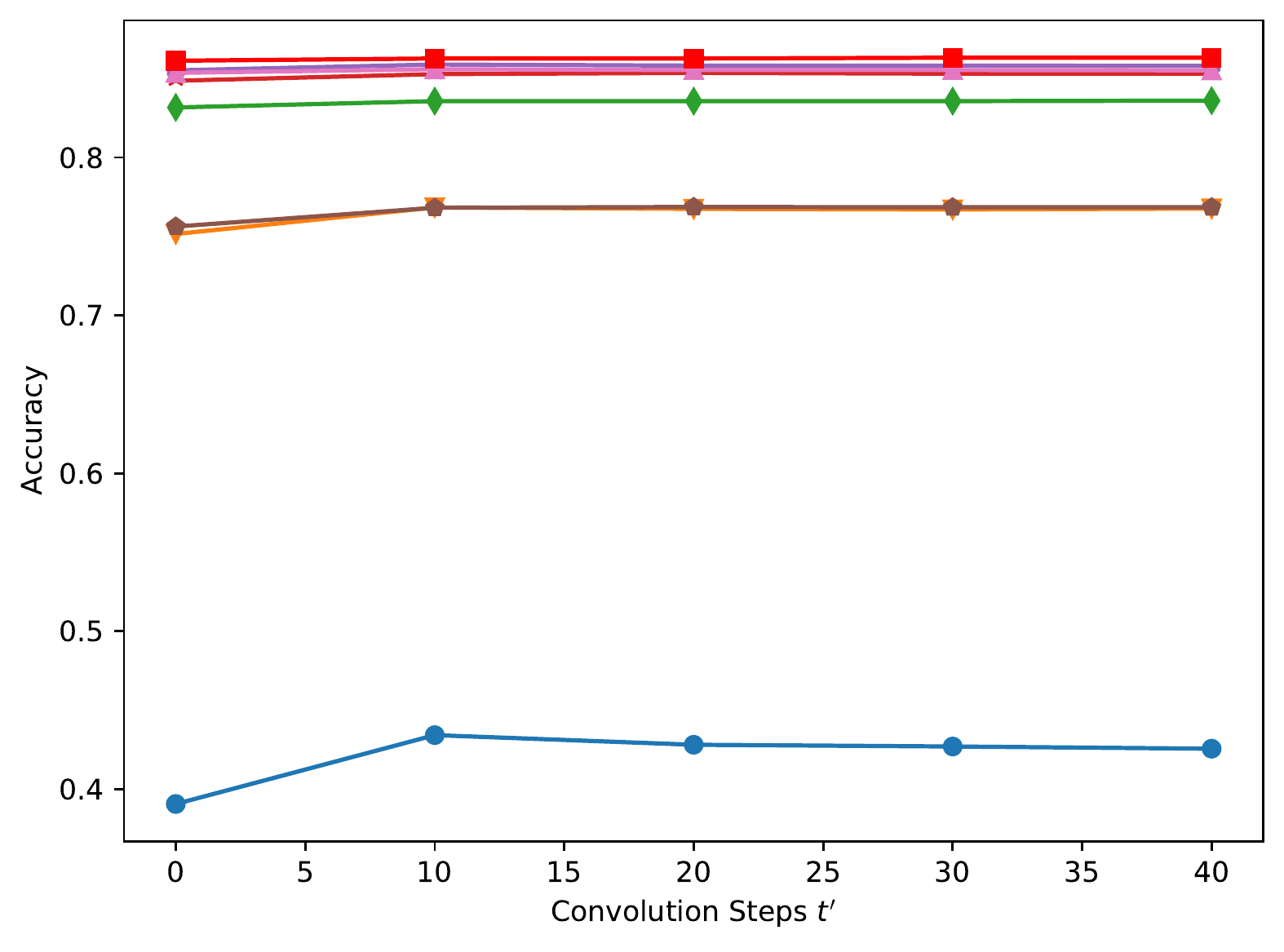}&
\includegraphics[width=0.45\linewidth]{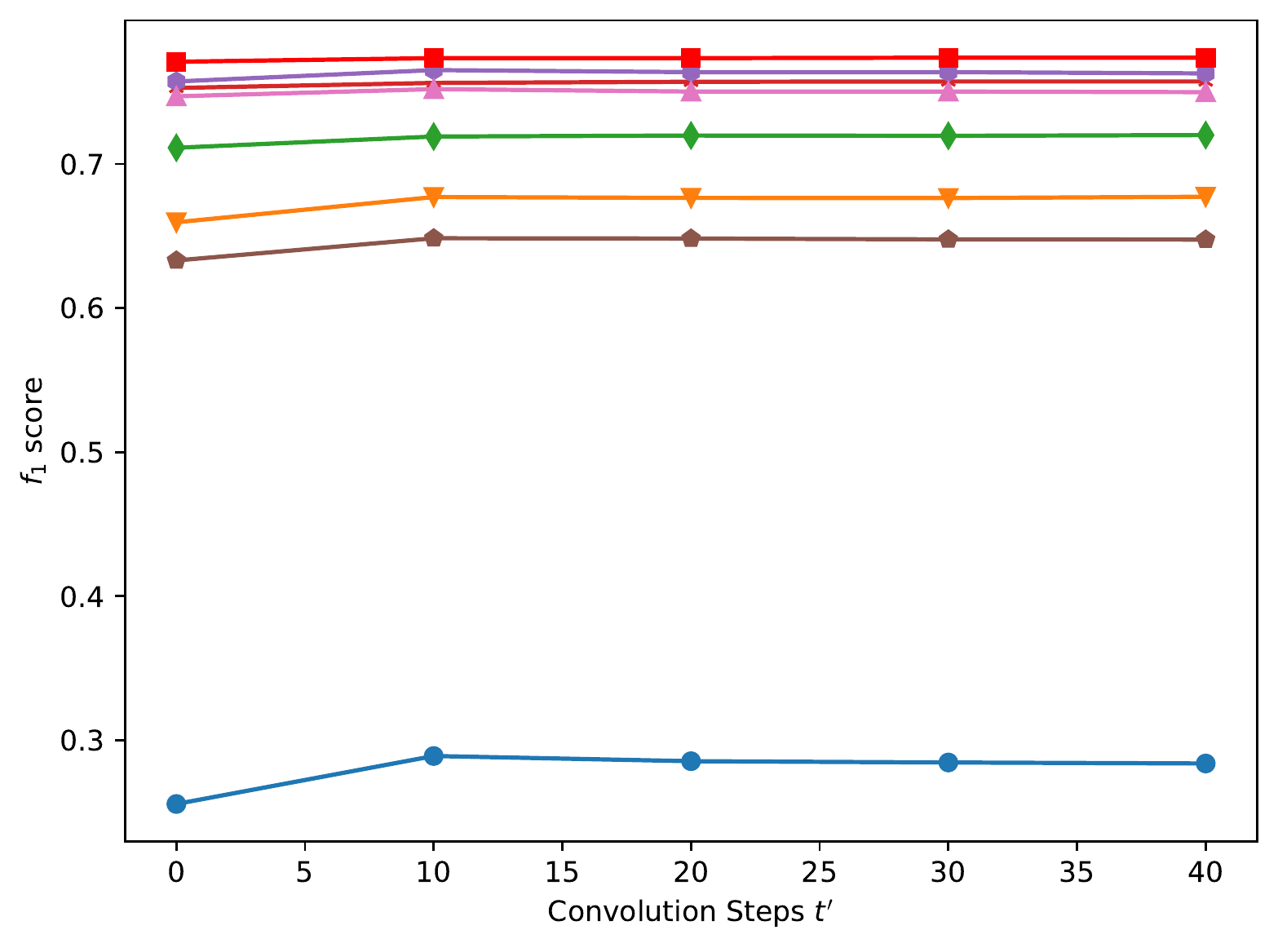}\\
\end{tabular}
\caption{Performance of the Compared Methods Attacked by SAP under Different Convolution Steps in Situation I.}
\label{fig:fig_1}
\end{figure*}

\begin{figure*}
\centering
\begin{tabular}{cc}
\multicolumn{2}{c}{\includegraphics[width=0.8\linewidth]{pics/1.pdf}} \\
\includegraphics[width=0.45\linewidth]{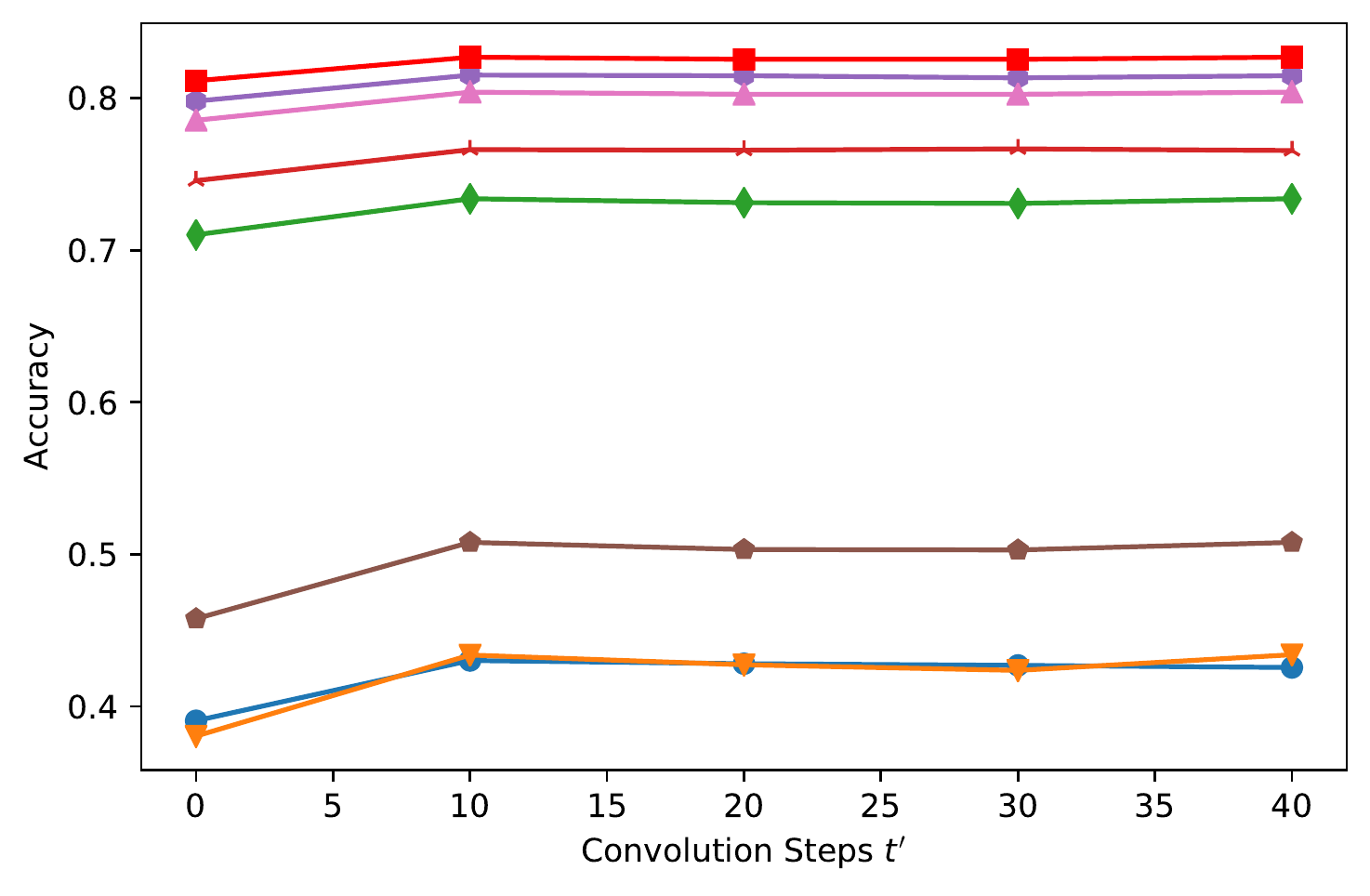}&
\includegraphics[width=0.45\linewidth]{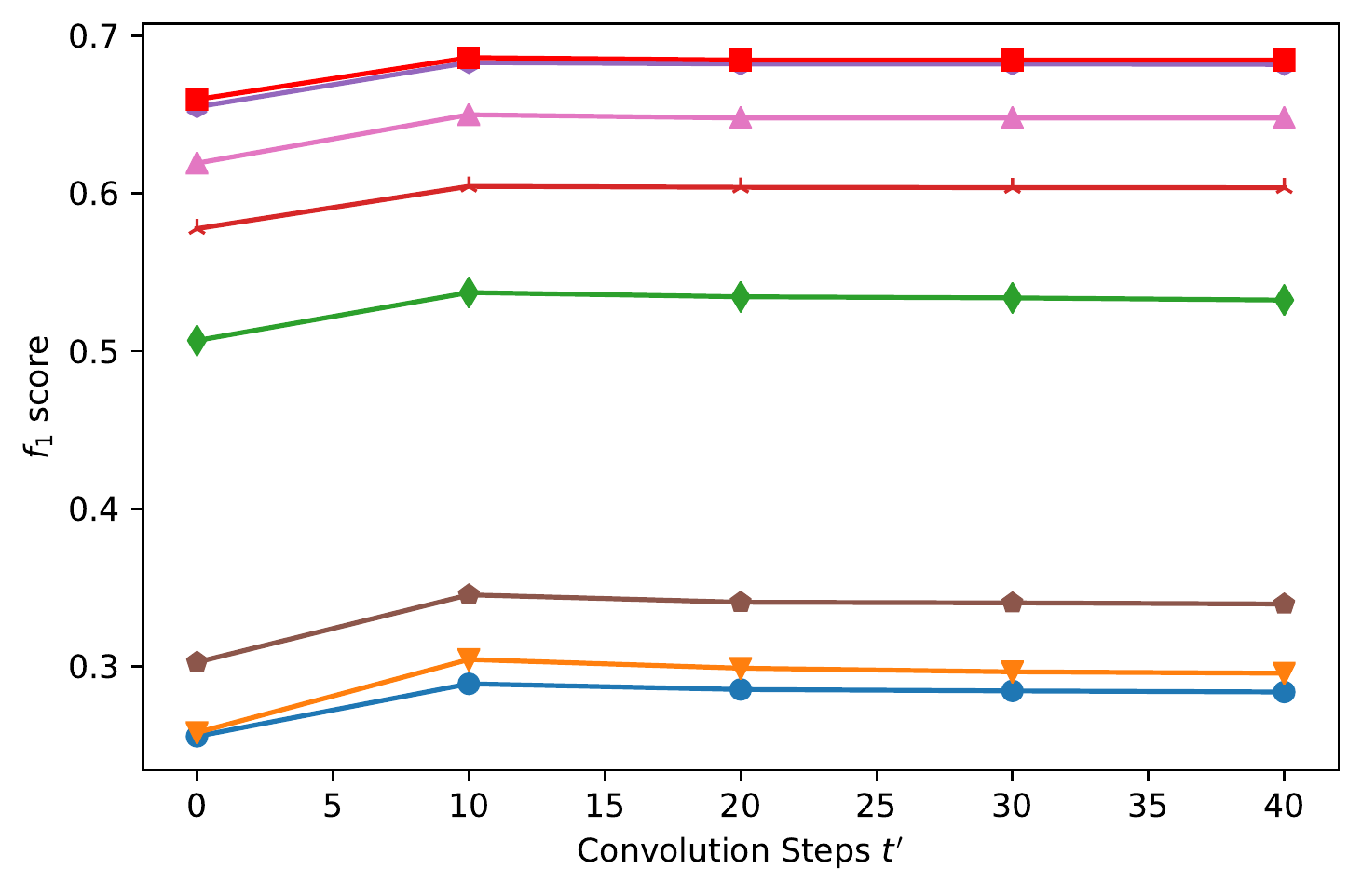}\\
\end{tabular}
\caption{Performance of the Compared Methods Attacked by SAP under Different Convolution Steps in Situation II.}
\label{fig:fig1}
\end{figure*}

To explore the defense effects of these defense methods under different-smoothing-degree SAP attacks, we changed the parameter $t'$ of the SAP attack, which controls the convolution times, and set the parameter $t'$ to $0,10,20,30,40$, respectively. The more convolution times, the smoother the attack noise becomes. As $t'$ takes $0$, SAP attack becomes PGD attack. The results in situation I and situation II are presented in \autoref{fig:fig_1} and \autoref{fig:fig1}, respectively. From these two figures, we can see that as the parameter $t'$ changes from $0$ to $10$, the accuracy ratio and F1-score of the classification model with explored defense methods are improved, and the increase is smaller in situation I, but larger in situation II. However, as $t'$ further increases, the accuracy ratio and F1-score do not change significantly, which means that the smoothness of the adversarial attack does affect the accuracy ratio and F1-score of the classification model with different defense methods. After $10$ convolutions, the smoothness of the adversarial attack is high and does not change significantly with more convolution operations, which leads that the accuracy ratio and F1-score of the classification model with different methods do not change significantly. Besides, we can see that the trained DNNs with ADT always had the best performance, and the order of defense effects for these defense methods remains the same under different convolution times.

\begin{table*}
\caption{Comparison of Methods under SAP attack in Situation I}\label{table:results_1}
\centering
\begin{tabular}{llll}
\toprule
\multicolumn{2}{l}{}                       & \multicolumn{1}{c}{Accuracy (Performance Drop)} & \multicolumn{1}{c}{$f_1$ score (Performance Drop)} \\ \midrule
                           & No Defense    & 0.4256$\pm$0.0727 (50.69\%$\pm$8.42\%)         & 0.2839$\pm$0.0656 (63.33\%$\pm$8.39\%)   \\ \hline
\multirow{4}{*}{Baselines} & JR            & 0.7676$\pm$0.0270 (11.03\%$\pm$3.16\%)         & 0.6772$\pm$0.0281 (12.69\%$\pm$3.06\%)   \\
                           & NSR           & 0.8359$\pm$0.0094 (3.39\%$\pm$0.71\%)         & 0.7200$\pm$0.0190 (5.75\%$\pm$1.62\%)   \\ 
                           & DD            & 0.7684$\pm$0.0110 (11.36\%$\pm$1.11\%)         & 0.6475$\pm$0.0128 (15.85\%$\pm$1.24\%)   \\
                           & AT            & 0.8551$\pm$0.0025 (1.03\%$\pm$0.38\%)         & 0.7498$\pm$0.0120 (2.11\%$\pm$1.70\%)   \\ \hline
\multirow{2}{*}{Variants}  & Init-ADT          & 0.8530$\pm$0.0044 (2.02\%$\pm$0.72\%)         & 0.7572$\pm$0.0114 (4.03\%$\pm$1.74\%)   \\
                           & Dist-ADT          & 0.8579$\pm$0.0082 (1.00\%$\pm$0.53\%)         & 0.7628$\pm$0.0135 (1.43\%$\pm$0.80\%)   \\ \hline
Proposed                   & ADT & \textbf{0.8631$\pm$0.0065 (0.67\%$\pm$0.43\%)}         & \textbf{0.7737$\pm$0.0145 (-1.02\%$\pm$1.84\%)}   \\
\bottomrule
\end{tabular}
\end{table*}

\begin{table*}
\caption{Comparison of Methods under SAP Attacks in Situation II}\label{table:results1}
\centering
\begin{tabular}{llll}
\toprule
\multicolumn{2}{l}{}                       & \multicolumn{1}{c}{Accuracy (Performance Drop)} & \multicolumn{1}{c}{$f_1$ score (Performance Drop)} \\ \midrule
                           & No Defense    & 0.4256$\pm$0.0727 (50.69\%$\pm$8.42\%)         & 0.2839$\pm$0.0656 (63.35\%$\pm$8.47\%)   \\ \hline
\multirow{4}{*}{Baselines} & JR            & 0.4223$\pm$0.0665 (51.08\%$\pm$7.70\%)         & 0.2958$\pm$0.0343 (61.82\%$\pm$4.43\%)   \\
                           & NSR           & 0.7339$\pm$0.0161 (15.43\%$\pm$1.89\%)         & 0.5323$\pm$0.0342 (31.28\%$\pm$4.42\%)   \\ 
                           & DD            & 0.5079$\pm$0.0246 (41.84\%$\pm$2.89\%)         & 0.3397$\pm$0.0249 (56.15\%$\pm$3.22\%)   \\
                           & AT            & 0.8040$\pm$0.0065 (7.01\%$\pm$0.67\%)         & 0.6477$\pm$0.0163 (16.39\%$\pm$2.10\%)   \\ \hline
\multirow{2}{*}{Variants}  & Init-ADT          & 0.7656$\pm$0.0114 (11.30\%$\pm$1.32\%)         & 0.6035$\pm$0.0193 (22.09\%$\pm$2.50\%)   \\
                           & Dist-ADT          & 0.8148$\pm$0.0082 (5.60\%$\pm$0.95\%)         & 0.6817$\pm$0.0101 (12.00\%$\pm$1.30\%)   \\ \hline
Proposed                   & ADT & \textbf{0.8270$\pm$0.0046 (4.35\%$\pm$0.36\%)}         & \textbf{0.6845$\pm$0.0127 (11.63\%$\pm$1.64\%)}   \\
\bottomrule
\end{tabular}
\end{table*}

\subsection{Defense Effects against PGD Attacks}

Similarly, we attack each trained DNN with or without defense method using the PGD at $\epsilon = 10, t = 20, t' = 0$. The mean value with the standard deviation of the prediction accuracy and the F1-score for each defense method under a PGD attack are shown in \autoref{table:results_2} and \autoref{table:results2}. From these tables, we can see that the accuracy and F1-score of the classification DNN with no defense are reduced to $39\%$ and $25\%$, respectively, and the decline of these two metrics is more than $50\%$. The classification DNN with ADT at $T=1$ still had the best performance, and its accuracy ratio against PGD attacks is still over $80\%$ in two situations, and its decline ratio of accuracy is lower than $1\%$ in situation I. In addition, we can see that adversarial training had good performance against low noise level PGD attacks, and NSR behaved better than JR under PGD attacks, which is consistent with the results of \cite{ma2020enhance}. Furthermore, the performance of the trained DNNs with defense methods in situation I is better that that in situation II, and the performance of the trained DNNs with ADT, Init-ADT, Dist-ADT, AT as well as NSR is more stable than that of the trained DNNs with JR and DD. In addition, the order of defense effects against low-noise PGD attacks is the same as that for SAP attacks in two different situations.

\begin{figure*}
\centering
\begin{tabular}{cc}
\multicolumn{2}{c}{\includegraphics[width=0.8\linewidth]{pics/1.pdf}} \\
\includegraphics[width=0.45\linewidth]{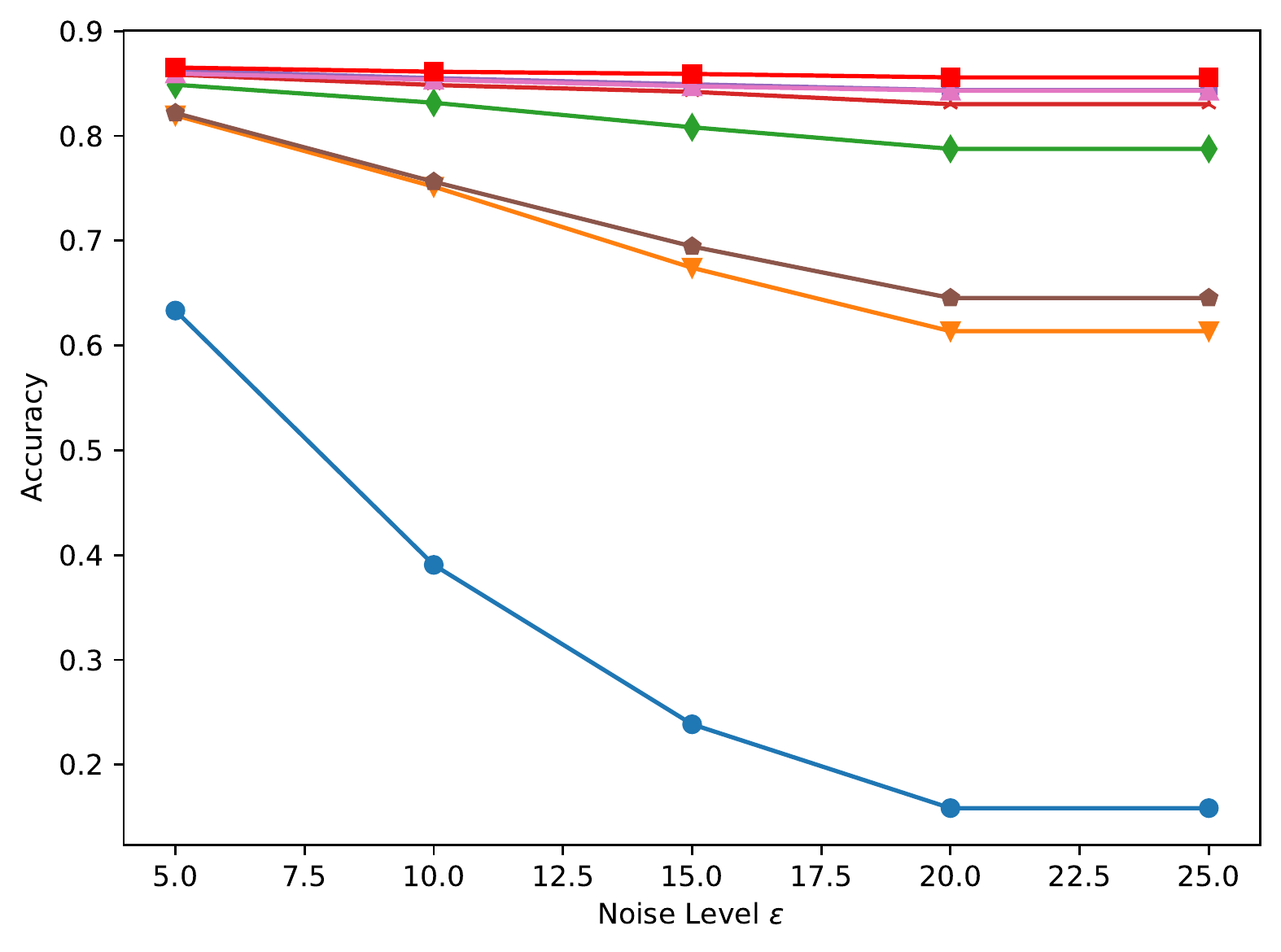}&
\includegraphics[width=0.45\linewidth]{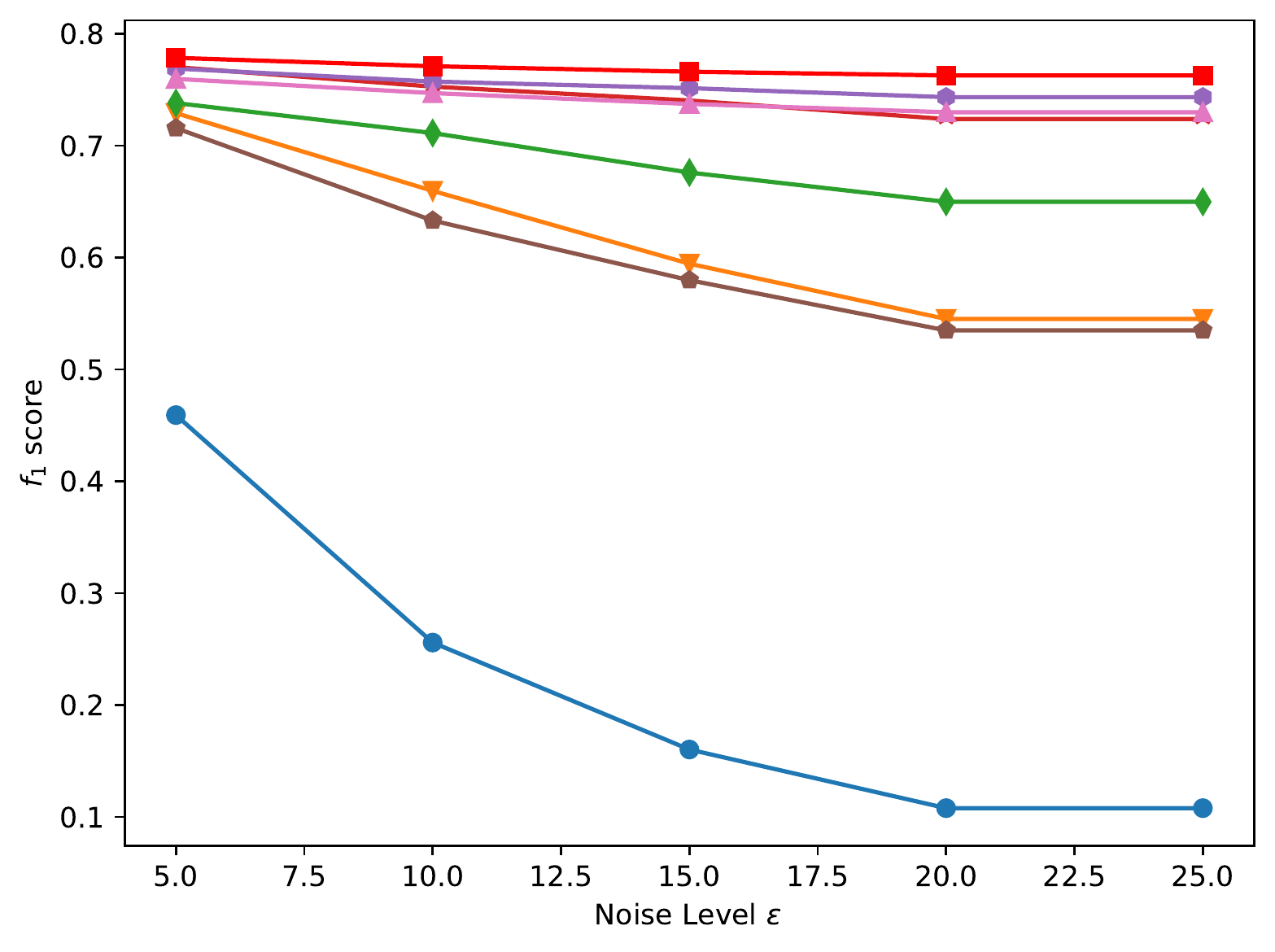}\\
\end{tabular}
\caption{Performance of the Compared Methods Attacked by PGD under Different Noise Levels in Situation I.}
\label{fig:fig_2}
\end{figure*}

\begin{figure*}
\centering
\begin{tabular}{cc}
\multicolumn{2}{c}{\includegraphics[width=0.8\linewidth]{pics/1.pdf}} \\
\includegraphics[width=0.45\linewidth]{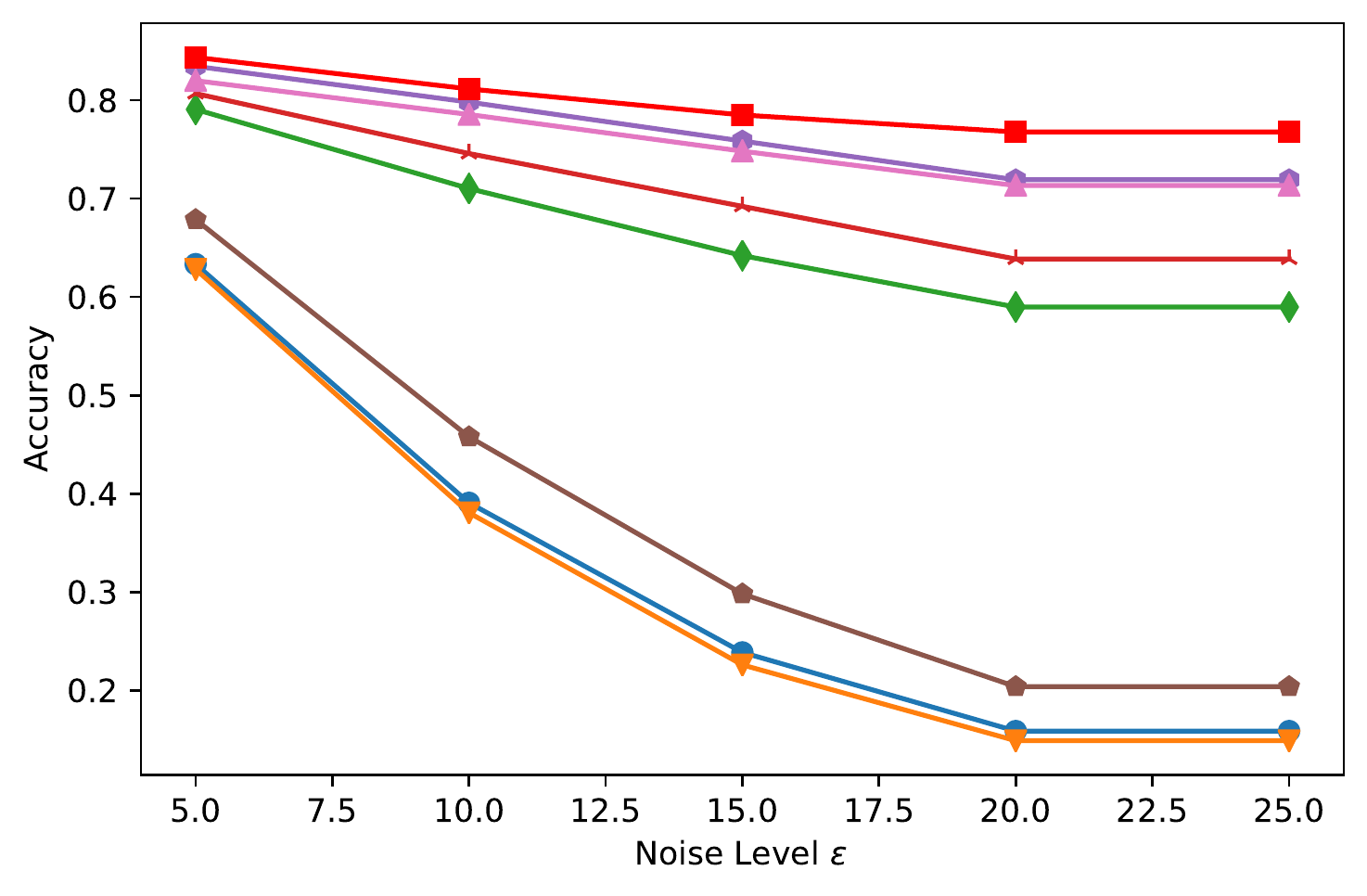}&
\includegraphics[width=0.45\linewidth]{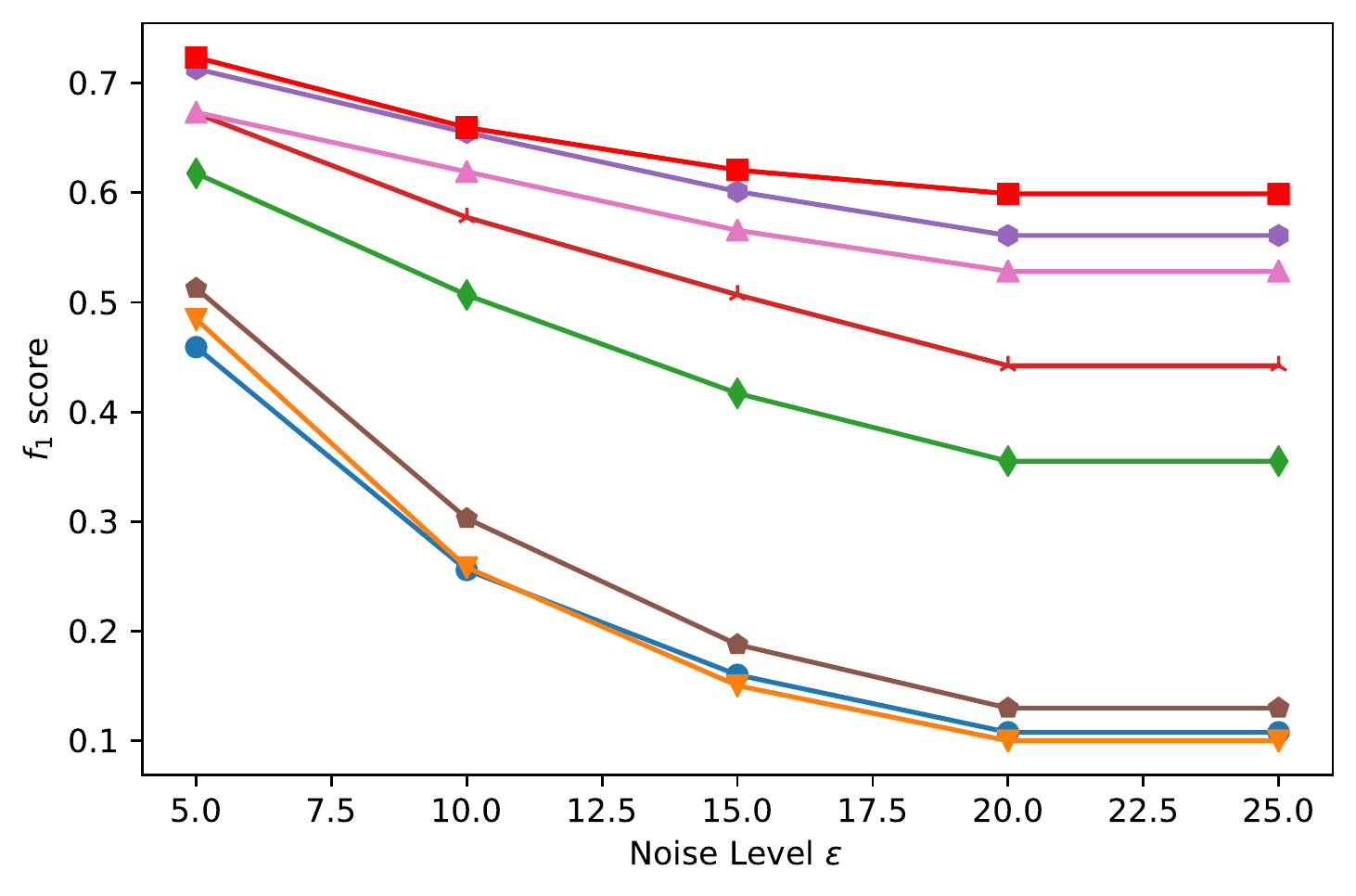}\\
\end{tabular}
\caption{Performance of the Compared Methods Attacked by PGD under Different Noise Levels in Situation II.}
\label{fig:fig2}
\end{figure*}

To further explore the performance of the explored defense methods under PGD attacks at different noise levels, we changed the parameter $\epsilon$. Specifically, we set $t'$ and $t$ as 0 and 20, respectively, and parameter $\epsilon$ as $5,10,15,20,25$, and the corresponding results for two situations are shown in \autoref{fig:fig_2} and \autoref{fig:fig2}. We can see that the accuracy ratio and F1-score of the trained DNNs with different defense methods decreased with the increase in noise level for PGD attack in the beginning, and remained fixed until the end of our experiments in two situations. Moreover, the decrease of accuracy ratio and F1-score for the trained DNNs with defense methods in situation I is lower than that in situation II. Furthermore, these figures show that the trained DNNs with ADT exhibited the best defense effects, indicating that although faced with PGD attacks within a certain low-level noise range, our method still has high defense effects. That is to say, our method has good robustness.

\begin{table*}
\caption{Comparison of Methods under PGD Attacks in Situation I}\label{table:results_2}
\centering
\begin{tabular}{llll}
\toprule
\multicolumn{2}{l}{}                       & \multicolumn{1}{c}{Accuracy (Performance Drop)} & \multicolumn{1}{c}{$f_1$ score (Performance Drop)} \\ \midrule
                           & No Defense    & 0.3906$\pm$0.0695 (54.80\%$\pm$8.08\%)         & 0.2558$\pm$0.0555 (66.94\%$\pm$7.16\%)   \\ \hline
\multirow{4}{*}{Baselines} & JR            & 0.7515$\pm$0.0272 (12.91\%$\pm$3.14\%)         & 0.6595$\pm$0.0276 (14.96\%$\pm$3.00\%)   \\
                           & NSR           & 0.8316$\pm$0.0118 (3.88\%$\pm$0.98\%)         & 0.7112$\pm$0.0232 (6.92\%$\pm$1.92\%)   \\
                           & DD            & 0.7562$\pm$0.0109 (12.76\%$\pm$1.09\%)         & 0.6330$\pm$0.0117 (17.73\%$\pm$1.06\%)   \\
                           & AT            & 0.8535$\pm$0.0030 (1.22\%$\pm$0.45\%)         & 0.7469$\pm$0.0153 (2.49\%$\pm$2.19\%)   \\ \hline
\multirow{2}{*}{Variants}  & Init-ADT          & 0.8485$\pm$0.0045 (2.53\%$\pm$0.76\%)         & 0.7527$\pm$0.0114 (4.59\%$\pm$1.83\%)   \\
                           & Dist-ADT          & 0.8551$\pm$0.0064 (1.33\%$\pm$0.43\%)         & 0.7572$\pm$0.0104 (2.15\%$\pm$0.54\%)   \\ \hline
Proposed                   & ADT & \textbf{0.8612$\pm$0.0050 (0.89\%$\pm$0.26\%)}         & \textbf{0.7709$\pm$0.0114 (-0.67\%$\pm$1.77\%)}   \\
\bottomrule
\end{tabular}
\end{table*}

\begin{table*}
\caption{Comparison of Methods under PGD Attacks in Situation II}\label{table:results2}
\centering
\begin{tabular}{llll}
\toprule
\multicolumn{2}{l}{}                       & \multicolumn{1}{c}{Accuracy (Performance Drop)} & \multicolumn{1}{c}{$f_1$ score (Performance Drop)} \\ \midrule
                           & No Defense    & 0.3906$\pm$0.0704 (54.74\%$\pm$8.16\%)         & 0.2558$\pm$0.0555 (66.97\%$\pm$7.17\%)   \\ \hline
\multirow{4}{*}{Baselines} & JR            & 0.3805$\pm$0.0756 (55.91\%$\pm$8.76\%)         & 0.2582$\pm$0.0462 (66.67\%$\pm$5.97\%)   \\
                           & NSR           & 0.7102$\pm$0.0158 (17.72\%$\pm$1.83\%)         & 0.5067$\pm$0.0270 (34.58\%$\pm$3.49\%)   \\
                           & DD            & 0.4577$\pm$0.0261 (46.97\%$\pm$3.02\%)         & 0.3028$\pm$0.0246 (60.90\%$\pm$3.18\%)   \\
                           & AT            & 0.7855$\pm$0.0039 (9.00\%$\pm$0.46\%)         & 0.6191$\pm$0.0157 (20.07\%$\pm$2.03\%)   \\ \hline
\multirow{2}{*}{Variants}  & Init-ADT          & 0.7458$\pm$0.0157 (13.59\%$\pm$1.82\%)         & 0.5776$\pm$0.0237 (25.44\%$\pm$3.06\%)   \\
                           & Dist-ADT          & 0.7981$\pm$0.0057 (7.53\%$\pm$0.66\%)         & 0.6548$\pm$0.0137 (15.46\%$\pm$1.77\%)   \\ \hline
Proposed                   & ADT & \textbf{0.8115$\pm$0.0036 (5.98\%$\pm$0.42\%)}         & \textbf{0.6595$\pm$0.0144 (14.86\%$\pm$1.86\%)}   \\
\bottomrule
\end{tabular}
\end{table*}

\subsection{Defense Effects against Boundary Attacks}

Without knowing the structure of the classification model, boundary attack, one of target black-box attack, generates adversarial samples by adjusting the samples of the target category to keep a small difference from the samples to be attacked. With these adversarial samples, hackers can make the classifier produce errors that meet their expectations. Therefore, in order to comprehensively explore the effectiveness of these defense methods, we did experiments to explore the defense effects of these methods against boundary attack.

Specifically, because we have trained $5$ classification DNNs without defense methods, we apply boundary attack to create adversarial samples based on these five DNNs and test data. Finally, we have created 291, 277, 300, 250 and 288 adversarial samples, respectively. Then, we use the trained DNNs with defense methods to classify these adversarial samples, and the results are showed as \autoref{table:results_3}. It should be noted that we only explore the results in situation I here, because the goal of boundary attack is to create adversarial samples identified as target categories without considering the inner structure of the classification model, and if we attack the trained DNNs with each defense method by boundary attack, the generated adversarial samples will not be identified correctly by themselves. We can also see that all trained DNNs with explored defense methods had good performance, and ADT also had excellent performance against boundary attack, which was close to the best one.

\begin{table}
\centering
\caption{Comparison of Methods under Boundary Attacks}\label{table:results_3}
\begin{tabular}{llll}
\toprule
\multicolumn{2}{l}{}                       & \multicolumn{1}{c}{Accuracy} & \multicolumn{1}{c}{$f_1$ score} \\ \midrule
\multirow{4}{*}{Baselines} & JR            & 0.8976$\pm$0.0298   & 0.8872$\pm$0.0375   \\
                           & NSR           & 0.8824$\pm$0.0257   & 0.8614$\pm$0.0398   \\
                           & DD            & 0.8877$\pm$0.0413   & 0.8702$\pm$0.0562   \\
                           & AT            & 0.9014$\pm$0.0211  & 0.8694$\pm$0.0413  \\ \hline
\multirow{2}{*}{Variants}  & Init-ADT          & \textbf{0.9163$\pm$0.0234}    & \textbf{0.9023$\pm$0.0342}   \\
                           & Dist-ADT          & 0.9005$\pm$0.0315   & 0.8840$\pm$0.0513   \\ \hline
Proposed                   & ADT & 0.9144$\pm$0.0238    &   0.8961$\pm$0.0304   \\
\bottomrule
\end{tabular}
\end{table}

\section{Discussions}
The results show that the classification model with ADT has a high accuracy ratio and F1-score under SAP attack and boundary attack, which are designed to attack an ECG classification network and whose created adversarial samples are not easily recognized by clinicians\cite{2020Deep}\cite{10.1145/3417312.3431827}. The defense effects of ADT against SAP attack and boundary attack are better than many traditional defense methods, such as JR, NSR regularization, adversarial training, and defense distillation in different situations. At the same time, ADT still performs well under low-noise PGD attacks, which have higher noise levels than SAP attacks. These phenomena show that our proposed model, ADT, has better defense effects and stronger robustness.

In addition, the results show that adversarial training has good defense effects against SAP attack, boundary attack as well as PGD attack of low-level noise. In the training process of adversarial training, the classification model needs to classify the adversarial samples created by SAP, and it will be punished by a loss function if it classifies the adversarial samples mistakenly. At the same time, compared with the original ECG samples, the morphology of the adversarial samples created by SAP did not change dramatically, so it was not difficult for the classification model with adversarial training to learn the characteristics of the SAP. Due to the punishment mechanism and the characteristics of SAP that are easy to learn, the classification model with adversarial training is robust against SAP as well as boundary attack whose generated adversarial samples don't change dramatically compared with the original samples.

Furthermore, defensive distillation has much better defense effects in situation I than that in situation II, which can be concluded from the corresponding results and denotes that the robustness of defensive distillation is not good. Adversarial samples created by SAP are added into the training processes of both network of ADT, the first network learns the morphological characteristics of the original ECG samples and adversarial samples, and then transmits this information to the second network of ADT, which improves the generalization ability of the classification model. In addition, the second network still learns the characteristics of nature samples and adversarial samples, which further enhances the generalization and robustness of the model. These are the reasons why ADT performs well, and from the truth that Dist-ADT performs better than Init-ADT in situation II of SAP attack and PGD attack, we can see that the latter plays a more important role in the performance of ADT.

\section{Conclusion}
In this study, we completely investigated the effects of defense methods against adversarial attacks targeting ECG classification deep neural networks. Furthermore, we propose a novel defense method called ADT, which involves adding adversarial samples into the training process of both networks of defensive distillation and is good at defending against adversarial attacks with small perturbations. The results of the experiments show that ADT has better defense effects against white-box attack including SAP attack and low-noise PGD attack which still have a higher level of noise than SAP, as well as black-box attack represented by boundary attack here.

In the future, we will explore the defense effects of gradient-free trained sign activation neural networks against SAP and evaluate more effective defense methods that require less training time but have better defense effects. In addition, we will also explore how to reduce the training time of the classification model with ADT, or achieve a small loss of defense effect but significantly reduce training time. We also plan to extend our work to obtain more explainable results. 

\section*{Acknowledgement}
This work was supported by the National Natural Science Foundation of China (No.62102008).

\bibliographystyle{unsrtnat}
\bibliography{references}  %%% Uncomment this line and comment out the ``thebibliography'' section below to use the external .bib file (using bibtex) .

%%% Uncomment this section and comment out the \bibliography{references} line above to use inline references.
% \begin{thebibliography}{1}

% 	\bibitem{kour2014real}
% 	George Kour and Raid Saabne.
% 	\newblock Real-time segmentation of on-line handwritten arabic script.
% 	\newblock In {\em Frontiers in Handwriting Recognition (ICFHR), 2014 14th
% 			International Conference on}, pages 417--422. IEEE, 2014.

% 	\bibitem{kour2014fast}
% 	George Kour and Raid Saabne.
% 	\newblock Fast classification of handwritten on-line arabic characters.
% 	\newblock In {\em Soft Computing and Pattern Recognition (SoCPaR), 2014 6th
% 			International Conference of}, pages 312--318. IEEE, 2014.

% 	\bibitem{hadash2018estimate}
% 	Guy Hadash, Einat Kermany, Boaz Carmeli, Ofer Lavi, George Kour, and Alon
% 	Jacovi.
% 	\newblock Estimate and replace: A novel approach to integrating deep neural
% 	networks with existing applications.
% 	\newblock {\em arXiv preprint arXiv:1804.09028}, 2018.

% \end{thebibliography}

\end{document}